\documentclass[structabstract]{aa}  
\usepackage{graphicx}
\usepackage{txfonts}
\begin{document}
   \title{Supersonic Cloud Collision - II}


   \author{S. Anathpindika
          \inst{1,2}
          }

   \institute{${}^{[1]}\textrm{I}$ndian Institute Of Astrophysics, 2$^{\textrm{nd}}$ Block - Koramangala, Bangalore 560 034; India \\ 
${}^{[2]}\textrm{S}$chool Of Astronomy \& Physics, Cardiff University, 5-The Parade, UK\\
              \email{${}^{[1]}\textrm{s}$umedh\_a@iiap.res.in \\
               ${}^{[2]}\textrm{S}$umedh.Anathpindika@astro.cf.ac.uk}
             }

   \date{Received September 00, 0000; accepted March 00, 0000}

 
  \abstract
   {}
   {In this second paper of the sequel of two papers, we further investigate the problem of molecular cloud (MC) collision. Anathpindika (2009) (hereafter paper I) considered highly supersonic cloud collisions and examined the effect of bending and shearing instabilities on the shocked gas slab. We now consider moderately supersonic cloud collisions (precollision cloud velocities of order 1.2 km s$^{-1}$ to 2.4 km s$^{-1}$).}
   {In the current paper, we present five SPH simulations of fast head-on and/or off-centre cloud collisions to study the evolution of ram pressure confined gas slabs. The relevant thermodynamics in the problem is simplified by adopting a simple barytropic equation of state. We explore the parameter space by varying the pre-collision velocity and the temperature of the post collision gas slab.}
   {The temperature in a pressure compressed gas slab is crucial to its dynamical evolution. The pressure confined gas slabs become Jeans unstable if the average sound crossing time, $t_{cr}$, of putative clumps condensing out of them, is much larger than their free fall time, $t_{ff}$. Self gravitating clumps may spawn multiple/larger $N$-body star clusters. Warmer gas slabs are unlikely to fragment and may end up as diffuse gas clouds. }
   {}


   \maketitle
%

\section{Introduction}

Giant Molecular Clouds (GMCs) have turbulent and clumpy interiors. Potential star forming clouds move randomly and have a velocity dispersion of the order of a few km s$^{-1}$ on the smallest scales (e.g. Larson 1981; Elmegreen 1997; Elmegreen 2000). One extreme of this problem was investigated in paper I, where we simulated highly supersonic, both, head-on and off-centre cloud collisions. We found that the dynamical behaviour of a shock compressed gas slab  was dominated by internal shear. Hydrodynamical instabilities suppressed the gravitational instability and turbulent mixing between slab layers made the slab susceptible  to the non-linear thin shell instability (NTSI) and the Kelvin-Helmholtz (KH) instability. Further, internal shear led to dissipation of internal energy via shocking between fluid layers. The shocked slab lost thermal support against self gravity and collapsed to form an elongated gas body, along the collision axis, quite similar to the integral filament in the Orion A region.

\emph{Star clusters and their birth} : Certain features of star formation are now clear. It has been observationally well established that  most star formation occurs in clusters (e.g. Pudritz 2002). There are two paradigms of star formation, one that of quiescent star formation - the cloud - core picture (Hoyle 1953; Whitworth \emph{et al} 1996) and second, the dynamical picture of star formation also called, triggered star formation.

The former picture is elucidated by self gravitating MCs. Potential star forming clouds may become self gravitating due to internal perturbations or may be compressed either by shock waves or galactic density waves.  Such theories are encouraged by spatial correlation between between star formation and the spiral arms of disk galaxies. Gravitational perturbations grow in regions which become Toomre unstable. Further, the surface density in star forming regions, $\Sigma_{gas}$, is supposedly related to the star formation rate (SFR), $\Sigma_{SFR}$, through the Schmidt-Kennicutt power-law (Schmidt 1959; Kennicutt 1989; Kennicutt, 1998).

However, this paradigm suffers on two counts. First, it inevitably leads to a correlation between the SFR and the density  wave amplitude. However, there is no corroborative observational evidence (e.g. Elmegreen \& Elmegreen 1986; Kennicutt 1989). Second, this paradigm fails to reconcile the observations of star formation in those spirals, which have small density wave amplitude (e.g. Block \emph{et al}  1994). This suggests  that the density waves and disk instabilities are not quintessential to star formation. We may therefore have to explore certain aspects of the second paradigm, stated above. Tan (2000) for instance, invoked cloud collision in the SFR calculations for a galactic disk and arrived at a  power-law similar to the Schmidt-Kennicutt correlation.

Triggered star formation is a self regulatory process. Rich, young star clusters have been observationally well studied. The orbital dynamics of clusters have enabled observers to estimate various physical properties of multiple systems like  orbital eccentricities and orbital stability, stellar masses, among other properties. Star clusters have reasonably homogeneous environs and therefore, are useful regions to study stellar evolution. For a  more detailed review of the matter refer (Clarke, Bonnell \& Hillenbrand 2000). Star clusters however, show a wide variation in stellar density. For instance, $\sim$96 percent of the IR sources in the Orion B region are distributed over four identified star forming regions. On the other hand, in the Orion A region, star forming cores are distributed along the north-south direction - the Integral shaped filament e.g. Lada \emph{et al} (1991); Meyer \& Lada (1999) and Mookerjea \emph{et al} (2000).

\emph{Birth of star clusters - Numerical studies }:  Triggered star formation is elucidated by dynamical interaction between fluid flows within  GMCs. Colliding flows produce gas slabs that are dynamically unstable (e.g. Klessen, Burkert \& Bate 1998; Klessen \& Burkert 2000). Dense gas shells swept up by expanding ionising  fronts from massive star clusters are also dynamically unstable and may fragment to produce a number of potential star forming clouds and filaments (e.g. Wada, Spaans \& Kim 2000; Dale, Bonnell \& Whitworth 2007; Furuya, Kitamura \& Shinnaga 2008; Lefloch, Cernicharo \& Pardo 2008).

The other model suggested and investigated further in the present work, is that of gravitational fragmentation of an external pressure confined gas slab. See for e.g. Elmegreen \& Elmegreen (1978),  Whitworth \emph{et al} (1994), Clarke (1999), and Boyd \& Whitworth (2005). Such slabs result from clouds colliding with relatively small precollision Mach numbers (e.g. Chapman \emph{et al} 1992; Bhattal \emph{et al} 1998).

 Below, we present five simulations, in that, two each are of fast head-on and off-centre cloud collisions, respectively. The remaining one is a slow head-on cloud collision which post collision, forms to a continuous shock as against a jump shock in the remaining four cases. We shall attempt to  contrast the findings recorded in paper I with those, presented here. The plan of the paper is as follows. In \S 2 below we briefly introduce the numerical scheme employed for our work and the code used. In \S 3 we describe the initial conditions and list the simulations performed. The results are discussed in \S 4 and we conclude in \S 5. The column density is measured in M$_{\odot}$pc$^{-2}$ and all the column density plots presented here, are plotted on  a logarithmic scale since they span about six orders of magnitude. The spatial coordinates are marked in parsecs.


\begin{table*}
\caption{List of simulations performed with relevant physical details.             }
\begin{center}
\begin{tabular}{c l c r c c  }     
\hline\hline       
 Serial & Experimental & Precollision & Number of  & Head-on & $T_{0}$\\
No. & details &  Mach Number($\mathcal{M}$) & particles & \\
\hline                    
   1 & $M_{cld1}$ = $M_{cld2}$ = 50 M$_{\odot}$ & 3 & $N_{gas}$ = 120000 & Yes & 54 K\\
& $R_{cld1}$ = $R_{cld2}$ = 0.8 pc, $T_{cld}$ = 54 K & &  \\
\hline
 2 & $M_{cld1}$ = $M_{cld2}$ = 50 M$_{\odot}$ & 3 & $N_{gas}$ = 120000 & No & 54 K\\
& $R_{cld1}$ = $R_{cld2}$ = 0.8 pc, $T_{cld}$ = 54 K & & \\
\hline
 3 & $M_{cld1}$ = $M_{cld2}$ = 50 M$_{\odot}$ & 3 & $N_{gas}$ = 120000 & Yes & 10 K\\
& $R_{cld1}$ = $R_{cld2}$ = 0.8 pc, $T_{cld}$ = 54 K & & \\
\hline
 4 & $M_{cld1}$ = $M_{cld2}$ = 2000 M$_{\odot}$ & 3 & $N_{gas}$ = 120000 & No & 10 K\\
& $R_{cld1}$ = $R_{cld2}$ = 4.5 pc, $T_{cld}$ = 377 K & & \\
\hline
 5 & $M_{cld1}$ = $M_{cld2}$ = 50 M$_{\odot}$ & 1 & $N_{gas}$ = 120000 & Yes & 54 K\\
& $R_{cld1}$ = $R_{cld2}$ = 0.8 pc, $T_{cld}$ = 54 K &&  \\
\hline                  
\end{tabular}
\end{center}
Note : The numbers on the left hand side below indicate the respective column numbers of the table.\\
(2) Physical details of individual clouds in a simulation (Mass, radius and temperature of individual clouds). \\
(3) Precollision Mach number ($\mathcal{M}$). \\
(4) Number of gas particles ($N_{gas}$) used in the simulations. \\
(5) Specifies whether the clouds collide head-on or at a finite impact parameter.\\
(6) Temperature at which the post shock gas slab  is maintained.
\end{table*}

   \begin{figure*}
   \vbox to 110mm{\vfil 
   \centering
       \includegraphics[angle=270, width=15.0cm]{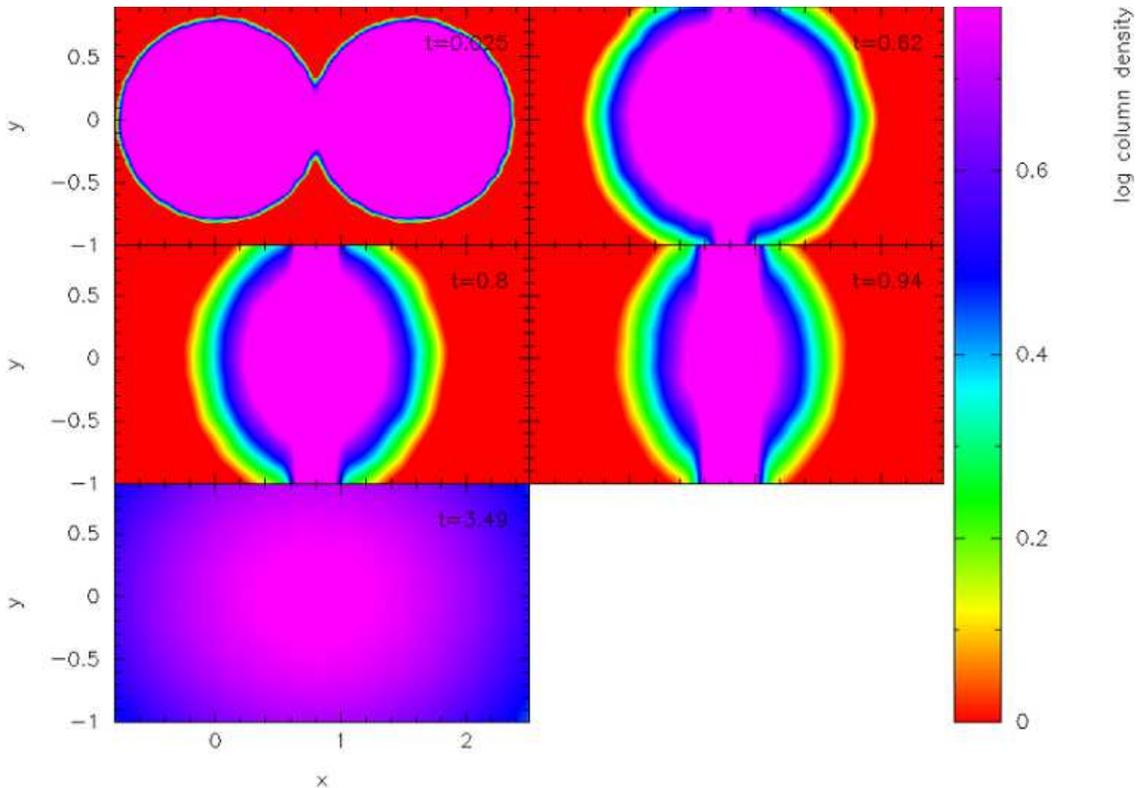}
   \caption{A column density plot of the time (measured in Myr) sequence of cloud collision in model 1. Cloud collision results in the formation of a pressure confined gas slab, which then re-expands and terminates in a diffuse gas cloud, as can be seen in the last snapshot ($t$ = 3.49 Myrs).}\vfil} \label{landfig}
    \end{figure*}
  \section[]{The Numerical Method \& Code}
  
We use the Smoothed Particle Hydrodynamics (SPH) numerical method for our simulations. This scheme treats the fluid under investigation as an ensemble of particles.  The simulations presented here are performed using an extensively tested SPH code, DRAGON (Goodwin, Whitworth \& Ward-Thompson 2004). It uses the Barnes-Hut Octal spatial tree (Barnes \& Hut 1986) to search nearest neighbours and, evaluate the net force on a SPH particle. The critical cell opening angle used by us is, $\theta_{crit}\sim$ 0.45 and each particle has 50$\pm$5 neighbours. In order to improve the accuracy of force calculations quadrupole moments of remote cells are also included. The code employs the multiple particle time stepping scheme (Makino 1991). 

An extremely dense agglomeration of particles in the simulation is replaced by a sink particle. The sink is characterised by two physical variables viz. the sink radius, $R_{sink}$, and the sink density, $\rho_{sink}$, both of which are predefined  (Bate, Bonnell \& Price 1995). We set $\rho_{sink}$ = 10$^{-12}$ g cm$^{-3}$ in all the simulations presented here. The sink radius is so chosen that the initial mass of a sink particle is comparable to the minimum resolvable mass, $M_{min}$, in the simulation. $M_{min}$ is essentially the same as the Bate-Burkert mass (Bate \& Burkert 1997). We had adopted an identical prescription in  paper I (refer \S 2 therein).  In the present work, protostellar cores are modelled as sinks.

\section[] {Initial Conditions \& Numerical Experiments }
   
We model MCs as  unconfined Bonnor-Ebert spheres. Thus, there is no external pressure on the cloud edges. Individual clouds are assembled in the same way as described in paper I. We compromise with the intercloud medium in these simulations to permit  larger number of gas particles in each cloud. 

In the first four models tested here, we adopt the same equation of state (EOS) as that in paper I albeit, with appropriate density switches. Using the temperature jump condition, we calculate the post shock temperature, $T_{ps}$, for each model. Behind the shock, the gas is allowed to cool. In the first two simulations, the gas downstream of the shock is maintained at original precollision cloud temperature, while in the next two, the gas is cooled down to 10 K.  The EOS employed is,
\begin{displaymath}
\frac{P(\rho)}{\rho} = (k_B/\bar{m})\times
\end{displaymath}
\begin{equation}
\left\{ \begin{array}{ll}
\Big(\frac{T_{cld}}{\textrm{K}}\Big) &; \rho\le10^{-21} \textrm{g cm}^{-3}\\
\Big(\frac{T_{cld}}{\textrm{K}}\Big)\Big(\frac{\rho}{\textrm{g cm}^{-3}}\Big)^{\gamma-1} &; 10^{-21}\textrm{g cm}^{-3} <\rho\le \\
& 2\times 10^{-21}\textrm{g cm}^{-3}\\
\Big(\frac{T_{ps}}{\textrm{K}}\Big) &; 2\times 10^{-21}\textrm{g cm}^{-3} <\rho\le\\&10^{-18}\textrm{g cm}^{-3}\\
\Big(\frac{T_{0}}{\textrm{K}}\Big)\Big[1 + \Big(\frac{\rho}{10^{-14}\textrm{g cm}^{-3}}\Big)^{\gamma-1}\Big]&; \rho > 10^{-18}\textrm{g cm}^{-3}.
\end{array} \right.
\end{equation}
Here $k_{B}$ and $\bar{m}$ are respectively, the Boltzmann constant and the average mass of a hydrogen atom, $T_{ps}$ is the post shock gas temperature. In cases 1 and 2, we set $T_{0}$ = $T_{cld}$, while in 3 and 4, $T_{0}$ = 10 K.

Finally, as a consequence of the very low precollision velocity ($\sim$ 0.4 km s$^{-1}$) of individual clouds in model 5, the post-collision shock is weak.  Weak shocks in astrophysical environs are nearly adiabatic, and so we model the one here, with a slightly stiffer EOS given by
\begin{equation}
\Big(\frac{T}{K}\Big) = \Big(\frac{T_{cld}}{K}\Big)\rho^{\gamma-1},
\end{equation}
where  all the symbols have their usual meanings and the adiabatic gas constant, $\gamma = \frac{5}{3}$ for either EOSs. Models tested here have been listed in Table 1 above along with  their relevant physical details. 

\begin{figure*}
 \vbox to 150mm{\vfil
 \includegraphics[angle=270, width=14cm]{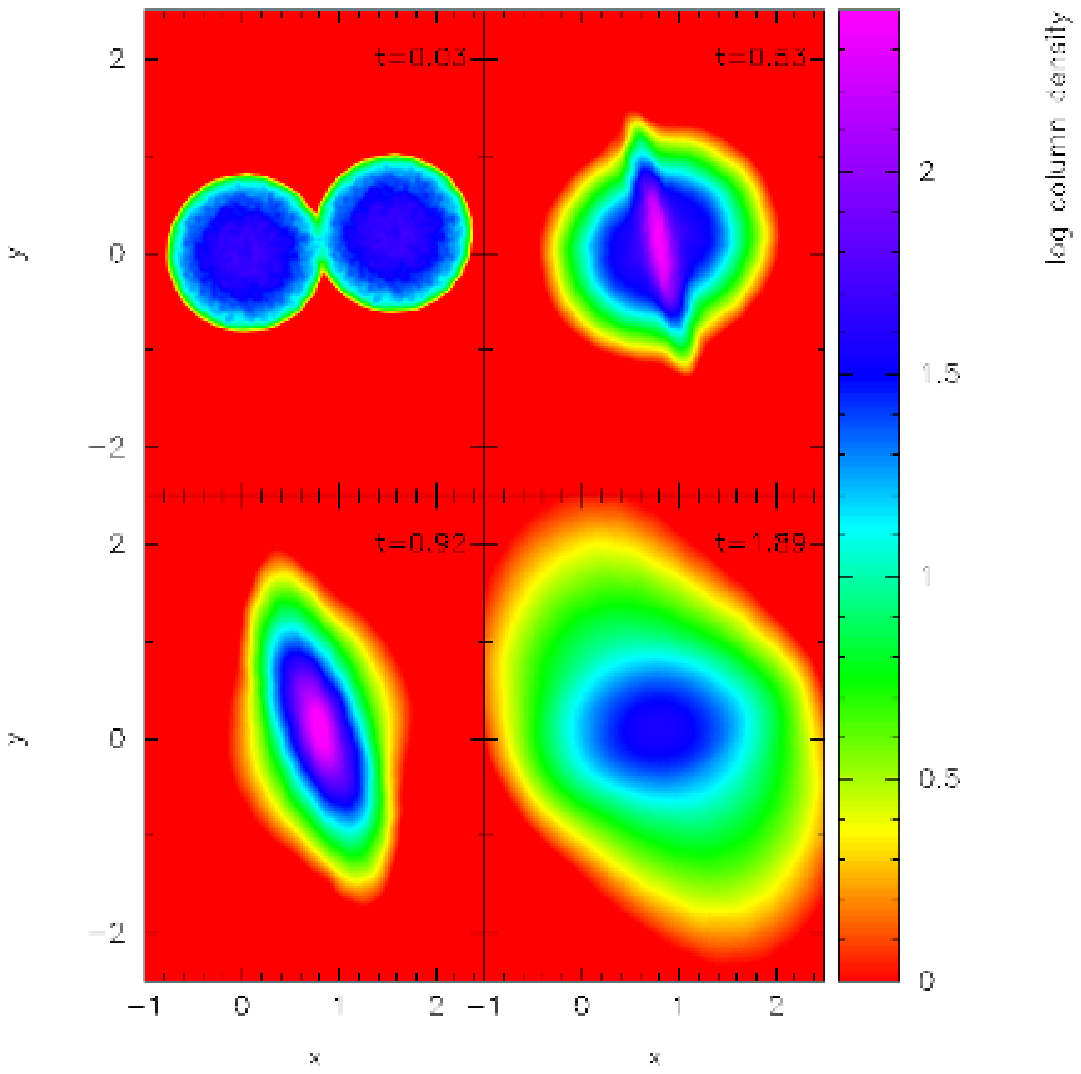}   
  \caption{Column density plot showing a time (measured in Myr) sequence tracking model 2. This sequence shows the colliding clouds, formation of the post-collision gas slab and then its re-expansion. } \vfil} \label{landfig} 
\end{figure*}

\section[]{Discussion }

The question of gravitational instability in cold, semi-infinite  planar gas slabs has been extensively studied in the past. However, most of these analyses were restricted to slabs with no external pressure on them. See for instance, McKee (1999) and Larson (1985) . On the other hand, stability analysis of cold, external pressure confined isothermal slabs by Elmegreen \& Elmegreen (1978) and Elmegreen (1989) suggests that the minimum mass of a fragment condensing out of such a slab, increases with the age of that slab as it continues to accrete matter. The monotonic growth time of the fastest growing unstable mode in such a slab is $\sim$0.25$(G\rho_{layer})^{-\frac{1}{2}}$ (Elmegreen 1989), where $\rho_{layer}$ is the average density of the post-shock gas slab.

Expressions for the length of the fastest growing mode, the timescale of growth of this mode and the minimum mass of the fragment condensing out of the slab have been derived by (Whitworth \emph{et al} 1994) and Boyd \& Whitworth (2005). Below we discuss the models tested in this work and compare the results of models 3 and 4 with the analytic findings of the latter authors.

\begin{figure*}
 \vbox to 150mm{\vfil
 \includegraphics[angle=270, width=14.0cm]{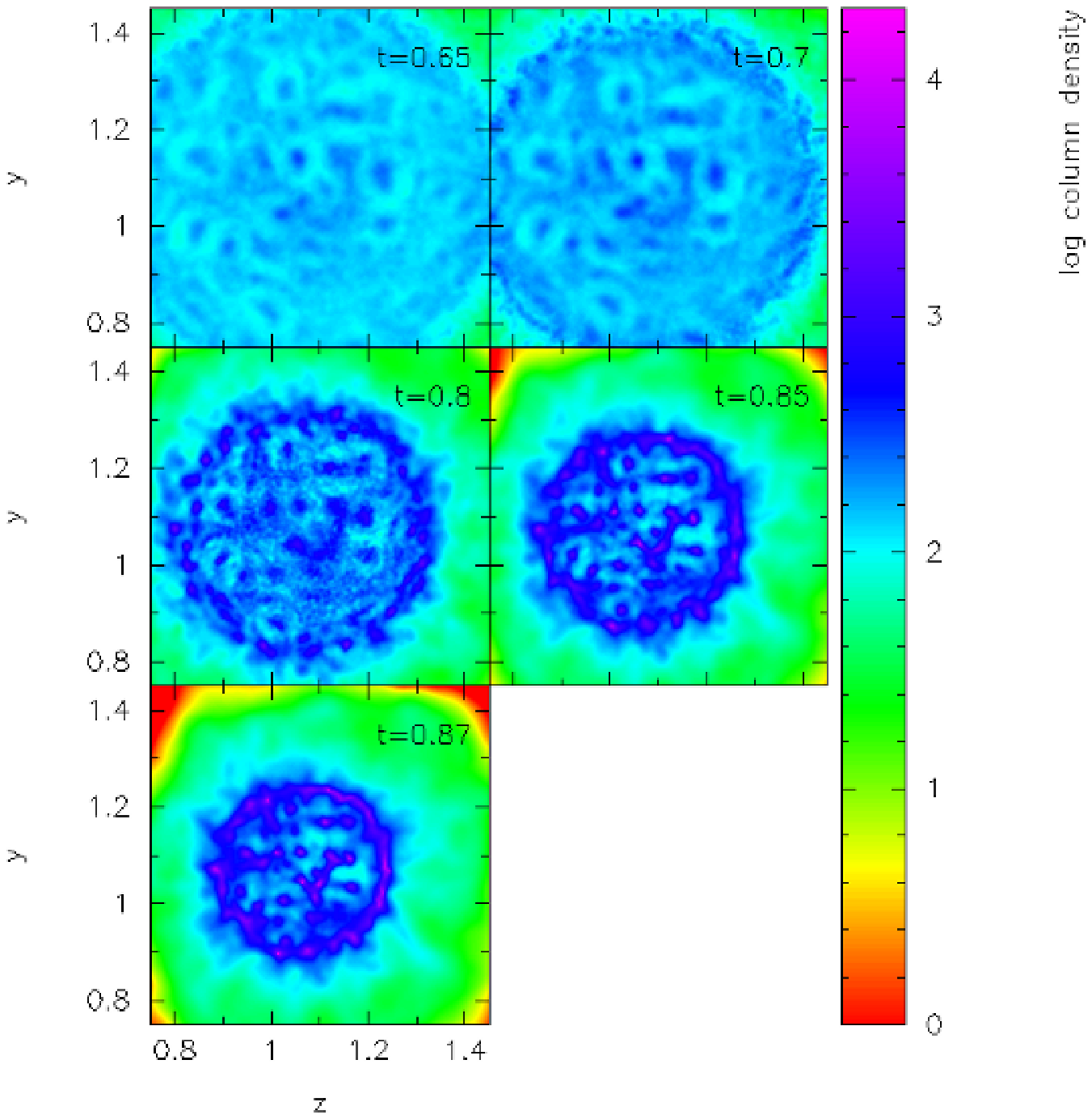}   
  \caption{A time (measured in Myr) sequence of column density plots showing the commencement of gravitational fragmentation,  then clumping and finally the fragmented planar gas slab in model 3.} \vfil} \label{landfig}
\end{figure*}

\begin{figure*}
 \vbox to 130mm{\vfil 
     \includegraphics[angle=270, width=14.0cm]{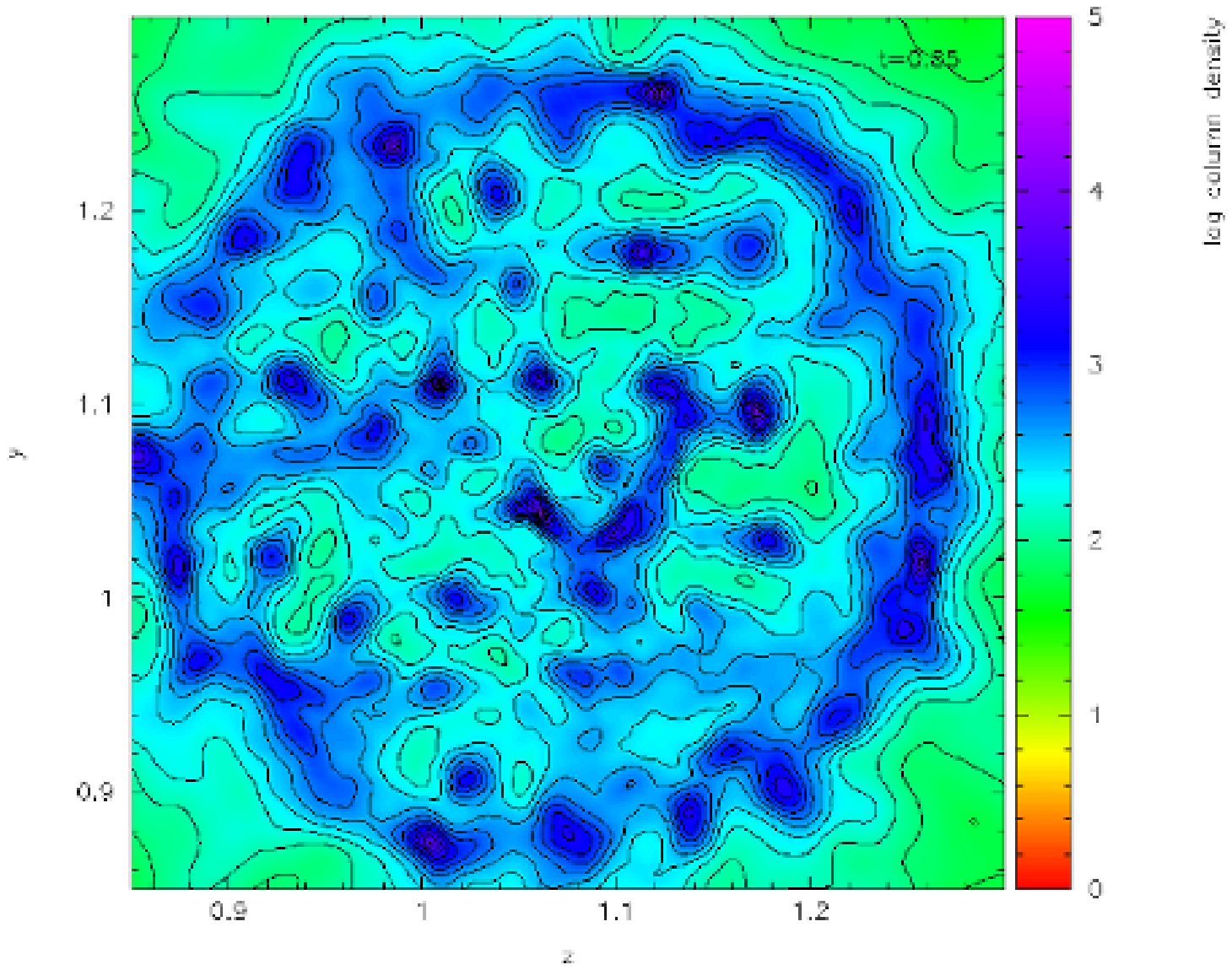}   
  \caption{A column density plot ($t$ = 0.85 Myr) of the fragmented slab in model 3. Density contours have been overlayed on it to facilitate identification of structure in it. A number of clumps and filaments can be seen in it. } \vfil} \label{landfig}
\end{figure*}

\textbf{Fast cloud collision ($T_{0}$ = $T_{cld}$; models 1 \& 2)}

In model 1, MCs of mass 50 M$_{\odot}$ each, collide head-on with each other at a pre-collision velocity of $\sim$1.3 km s$^{-1}$ ($\mathcal{M}$ = 3). Individual clouds are thermally supported against self gravity. The post-collision thermodynamics in this model is governed by the EOS defined by equation (1) above. The clouds collide and form a planar gas slab confined by  ram pressure. The time sequence of colliding clouds and their evolution after collision is  shown in Fig.1, which is  a column density plot showing the clouds and the post-collision slab.  We treat the gas slab as approximately isothermal and hold it  at the original pre-collision cloud temperature.  Our assumption of the post collision temperature should not have any adverse bearing on the final outcome of the simulation, since the post-shock temperature, $T_{ps}$, calculated using the temperature Jump condition will only be higher, making the slab even warmer.

The post-shock gas slab in this model does not undergo gravitational fragmentation. After formation, it simply re-expands and ends up as a diffuse gas cloud ( $t$ = 3.49 Myrs in Fig. 1). Unlike the gas slabs reported in paper I, the one in this model does not show any evidence for the bending instability. However, gas elements within the slab are non-static and dissipate mechanical energy via shocking, facilitated by the artificial viscosity. The gas slab in this case generally evolves via an interplay between gravity and thermal pressure. There are two phases in this evolution. In the first, thermal pressure builds up in the slab after shocking and then, in the next phase, as we permit the gas to cool, thermal pressure within the slab rapidly diminishes, and the slab re-expands. Even then the slab is too warm to  become Jeans unstable. This seems similar to the re-expansion of the shocked slabs reported in paper I. However, unlike those slabs, the one in this case does not collapse to form an elongated object along the collision axis suggesting that, gravity and thermal pressure attain a dynamic equilibrium.

The gas slab in this case is not sufficiently massive, or in other words, it is not sufficiently cold (for the mass that it has) so as to support the Jeans instability. According to Whitworth \emph{et al} (1994), the length of the fastest growing unstable mode is
\begin{equation}
L_{fastest} \sim \frac{2a_{layer}^{2}}{G\Sigma_{layer}},
\end{equation}
which is of order a few parsecs for the present case, at least an order of magnitude  larger than the slab thickness. 

In model 2, we maintain  physical conditions identical to those in the previous case except, that the clouds now collide, off-centre. The impact parameter, $b$ , is a fourth of the cloud radius, $R_{cld}$. The  time sequence of colliding clouds in this model is shown in Fig. 2. The cloud collision results in an oblique shocked slab unlike a planar slab in model 1. After formation, the slab tumbles about the $z$-axis and eventually re-expands, as in model 1. This can be seen in the snapshot corresponding to $t$ = 1.89 Myrs in Fig. 2. 

As in model 1, the gas slab in this case also evolves through an interplay between gravity and thermal pressure. Contrary to intuition, the slabs in either cases evolve on a similar time timescale.  This suggests that the re-expansion of the oblique slab also commences at a similar epoch as the planar slab. In fact, the oblique slab expands slightly faster than the planar slab of model 1. This is presumably due to axial transfer of angular momentum as the outer ends of the oblique slab move outward (see snapshot corresponding to $t$ = 1.89 Myrs in Fig. 2). The terminal state in this case is similar to that in case 1.

Equation (3) above, is valid for off-centre cloud collision as we have chosen a rather small impact parameter. $L_{fastest}$, calculated using this equation in the present case is much larger than the slab thickness. Hence as in model 1, the gas slab in this case as well, does not become Jeans unstable. \emph{Such a scenario (as in models 1 and 2) will thus, simply lead to formation of diffuse clouds in the ISM. These clouds may spawn star formation only when they have been sufficiently squeezed.}

\begin{figure}
\centering
 \vbox to 80mm{\vfil 
     \includegraphics[angle=0, width=8.5cm]{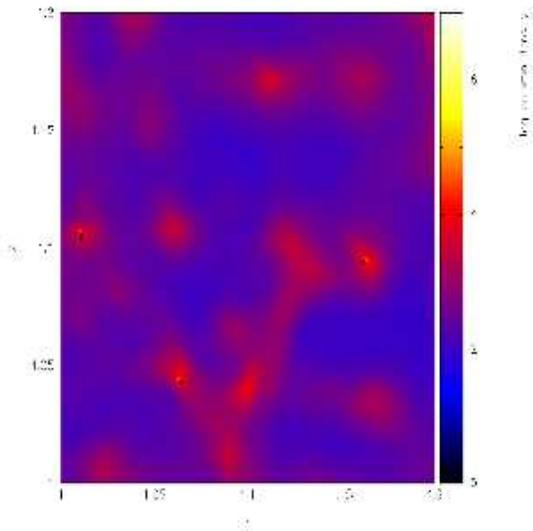}   
  \caption{A column density plot showing a closer view of the central region of the fragmented slab in model 3. The (three) black dots are the (three) sinks formed in various clumps. } \vfil} \label{landfig}
\end{figure}

\textbf{Fast cloud collision ($T_{0}$ = 10 K; models 3 \& 4)}

We now repeat the simulations preformed in models 1 and 2 in our next two models, 3 and 4 respectively, but with one change. We now hold the post shock gas slab at 10 K. As discussed above, we intend to explore the effect of cooling on the dynamical evolution of the gas slab. We saw in models 1 and 2 above, that the gas slab at 54 K was not susceptible to the gravitational instability. 

However, in model 3, where the MCs collide head-on, we observe that the post-collision gas slab undergoes gravitational fragmentation. In Fig. 3, we present a time sequence of the gas slab after its formation, as seen face-on. As the gravitational instability starts growing, the slab shows signs of flocculation ($t$ = 0.65 Myr, $t$ = 0.7 Myr in Fig. 3), and eventually clumps in the slab condense out.  Some of the smaller clumps merge to form larger clumps while some others lie embedded in filamentary structures. Most of the clumps being spheroidal, appear non-circular in the projection maps (see Fig. 4 below).

The gravitational instability grows on a time scale much smaller than  the free fall time of the individual precollision clouds, or even the cloud crushing time, $t_{cr}$ \footnote{$t_{cr} = \frac{2R_{cld}}{v}$, where $v$ is the precollision speed of a cloud and $R_{cld}$ is its radius.}. We now compare observations from this model with corresponding analytic predictions.  In our simulation we observe that the clump formation commences after about 0.65 Myr. This is the time required for the growth of unstable mode, $t_{growth}$, leading to condensation of a clump and defined as
\begin{equation}
t_{growth} \sim \Big(\frac{L}{G\sigma_{layer} - a^{2}L^{-1}}\Big)^{\frac{1}{2}},
\end{equation}
Whitworth \emph{et al} (1994).

Here $L$, $\sigma_{layer}$, and, $a_{layer}$, respectively are the size of the clump, average surface density of the slab, and the average  sound speed in the slab. Plugging in the appropriate values from the simulation we get $t_{growth}$ = 0.6 Myr, which is in agreement with the epoch observed in the simulation. We again note that, clump formation commences after about 0.65 Myr. Thereafter the gravitational instability grows quickly and the slab fragments into a number of clumps. The minimum mass of a fragment is given by
\begin{equation}
M_{frag} = \mathcal{M}^{\frac{1}{2}}\frac{a_{layer}^{3}}{(G^{3}\bar{\rho}_{layer})^{\frac{1}{2}}},
\end{equation}
 Whitworth \emph {et al} (1994). Here $\bar{\rho}_{layer}$ is the average density of the layer and $\mathcal{M}$ is the precollision Mach number. Using equation (5), we calculate the mass of this fragment, $M_{frag}$, as 0.32 M$_{\odot}$ (0.22 M$_{\odot}$) for the present simulation. The number in brackets is the minimum mass of the fragment that condenses out of the slab. 

The fragmentation of this slab results in the formation of 36 big and small clumps. Fig. 4 shows a close-up of the fragmented slab. At this epoch a few clumps have already become self gravitating while a few others, oscillate for some time before collapsing under self gravity. The column density plot in Fig. 5 shows a closer view of the central region of this slab. Three sinks have been marked by black dots in as many clumps.  Yet some other small clumps merge with similar clumps and form larger clumps and filaments.  To facilitate identification of structure in the slab, we have overlayed density contours on the column density plot in this figure.

\begin{figure*}
\centering
  \vbox to 220mm{\vfil
     \includegraphics[angle=270]{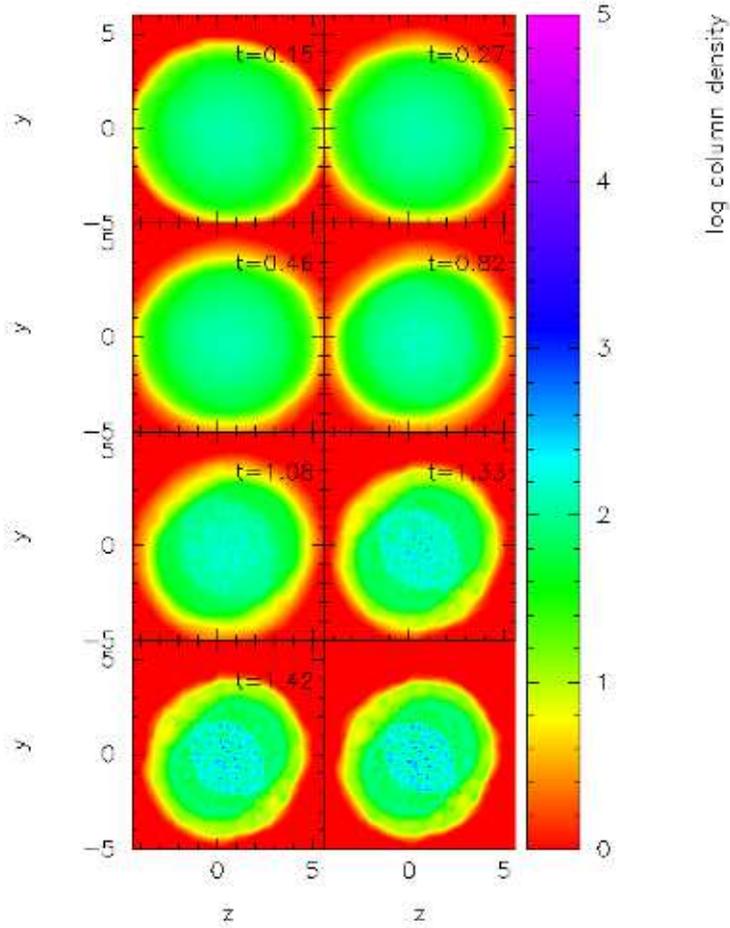}   
  \caption{A time (measured in Myr) sequence of column density plots of the oblique slab in model 4. A face-on view of the slab has been shown in these plots.  We can see that as the slab grows old, it shows signs of flocculation ($t$ = 0.82 Myr) and subsequently fragments (right hand plot in the third row, $t$ = 1.6 Myrs). The last plot shows a network of filaments and clumps embedded in them.}\vfil} \label{landfig}
\end{figure*}

The slab as seen in Figs. 4 and 5 is fairly evolved and at this epoch, the gravitational instability has fully grown. Almost all of the gas in the slab ends up in density structures. The average density of the structure in the slab is $\sim$10$^{-18}$ g cm$^{-3}$, which then suggests that the growth period of the instability as defined by Elmegreen (1989), and quoted above, is $\sim$ 0.8 Myr. This is comparable to the epoch (0.85 Myr) by which the instability has fully grown.

According  to Nakamura, Hanawa \& Nakano (1993), the characteristic length scale of a filamentary structure is 
\begin{equation}
H = 0.0035\Big(\frac{a_{layer}}{0.3 \textrm{km s}^{-1}}\Big)\Big(\frac{\bar{n}_{layer}}{2\times 10^{6} \textrm{cm}^{-3}}\Big)^{-\frac{1}{2}} \textrm{pc},
\end{equation}
where all the symbols have their usual meanings. The average temperature of the densest regions in the slab is about 10K  while $\bar{n}_{layer}\sim 10^{7}$ cm$^{-3}$. Substituting these values in equation (6) above, we get $H\sim$ 0.13 pc. The gravitational fragmentation of the filaments produce clumps and the separation between individual clumps is of the order of Jeans length, $\lambda_{J}$, which for a filament is $\lambda_{J}\sim 22H$ (Nakamura, Hanawa \& Nakano 1993). In the present case, $\lambda_{J} \sim$ 0.3 pc. This agrees with the average separation between clumps embedded in filamentary structures, as seen in  Fig. (3) above. This simulation was terminated after $t\sim$ 0.87 Myr by which time seven sinks had formed in various clumps.

Next, in model 4  the cloud collision produces an oblique gas slab confined by ram pressure. Like in the previous case (model 3), this slab also undergoes gravitational fragmentation. However, the fragmentation observed here is somewhat different from that reported by Chapman \emph{et al} (1992) and Pongracic \emph{et al} (1992). According to these authors, the oblique gas slab tumbles about a direction perpendicular to the plane of collision and breaks in to two blobs, which then undergoes secondary fragmentation to form multiples. Such a fragmentation resembles the bar mode instability. In the present case, although the oblique slab exhibits a similar tumbling motion the final outcome is different from that reported by the respective authors. 

A possible reason for this observed difference could be, a smaller Jeans length in the present case. Our pre-collision clouds being more massive produce a slab with greater column density, which shortens the Jeans length and enhances its growth rate. In this eventuality, the gravitational instability will dominate over other unstable modes of the tumbling slab, notably the bar mode. The fragmentation of the slab after its formation can be seen in Fig. 6. This is a time sequence of column density plots of the gas slab, as seen face-on. The nature of fragmentation observed here is in consonance with the predictions of Whitworth \emph{et al} (1994) for an off-centre cloud collision, with a small impact parameter. 

As in the previous case, we calculated the dynamical properties for this slab, using equations (3), (4) and (5). The length of the fastest growing unstable mode in the slab, $L_{fast}$, defined by equation (3) above is, indeed much smaller than the radial extent of the slab. Clumping in the slab commences after $\sim$ 0.4 Myr, while the smallest clump has mass $\sim$ 12 M$_{\odot}$ and size $\sim$ 0.1 pc. Corresponding analytic calculations yield 0.4 Myr, 11 M$_{\odot}$ and 0.12 pc, respectively. The calculated values thus, seem to be in good agreement with those observed in the simulation. 

Once the slab becomes unstable, it develops floccules and soon fragments. The final state of the slab, post-fragmentation, can be seen in Fig. 7 which is a column density plot of the slab as seen face-on. The clumps and filaments in it have  density of order 10$^{-19}$ g cm$^{-3}$. Density contours overlayed on this plot elucidate the structure in it. The average smoothing length of SPH particles in these simulations is at least an order of magnitude smaller than $L_{fast}$. We therefore think that the observed fragmentation of gas slabs is likely to be physical.

\emph{We find that fragmentation of a cold gas slab leads to clump formation and self gravitating clumps may spawn star clusters, a view that as also been advanced by numerous workers before.} See for instance Chapman \emph{et al} (1992), Bhattal \emph{et al} (1994) and Clarke (1999), among others. Here we can see that fragmenting gas slabs also produce dense filaments. In fact Burkert \& Hartmann (2004) have performed 2-d grid simulations of self gravitating, finite gas sheets having different geometries. It makes an interesting comparison between two of their models viz. the 'static circular sheet' and the 'rotating elliptical sheet', with the respective gas slabs in models 3 and 4 discussed in this section.

While the formation of the outer ring like structure in the gas slab of model 3 (see Fig. 4), looks similar to the structure in their collapsing circular sheet, but unlike the global collapse observed in their model the gas slab discussed here undergoes gravitational fragmentation. However, the reason behind formation of this ring, that of edge material initially piling up, might still be valid. Similarly, the Burkert \& Hartmann (2004) model of collapsing elliptical gas sheet leading to the formation of a filament is at variance with the result of model 4, discussed here. This is obviously due to the faster growth rate of the gravitational instability, as compared to the global collapse time of the gas slab. However, the shocked slab (both planar and oblique), where the gravitational instability is apparently suppressed by hydrodynamical instabilities, undergoes  global collapse (see paper I). Hartmann \& Burkert (2007) have proposed a rotating elliptical sheet model to explain the formation of the Orion Integral Filament.

Observational evidence for similar occurrences have been reported, for instance by Lada, Alves \& Lada (1999) and Elmegreen (2002) about a star cluster embedded in an elongated region, IC546. This cluster is thought to have been formed due to some dynamical trigger  
  \begin{figure*}
  \centering
 \vbox to 130mm{\vfil 
     \includegraphics[angle=0,width=15cm]{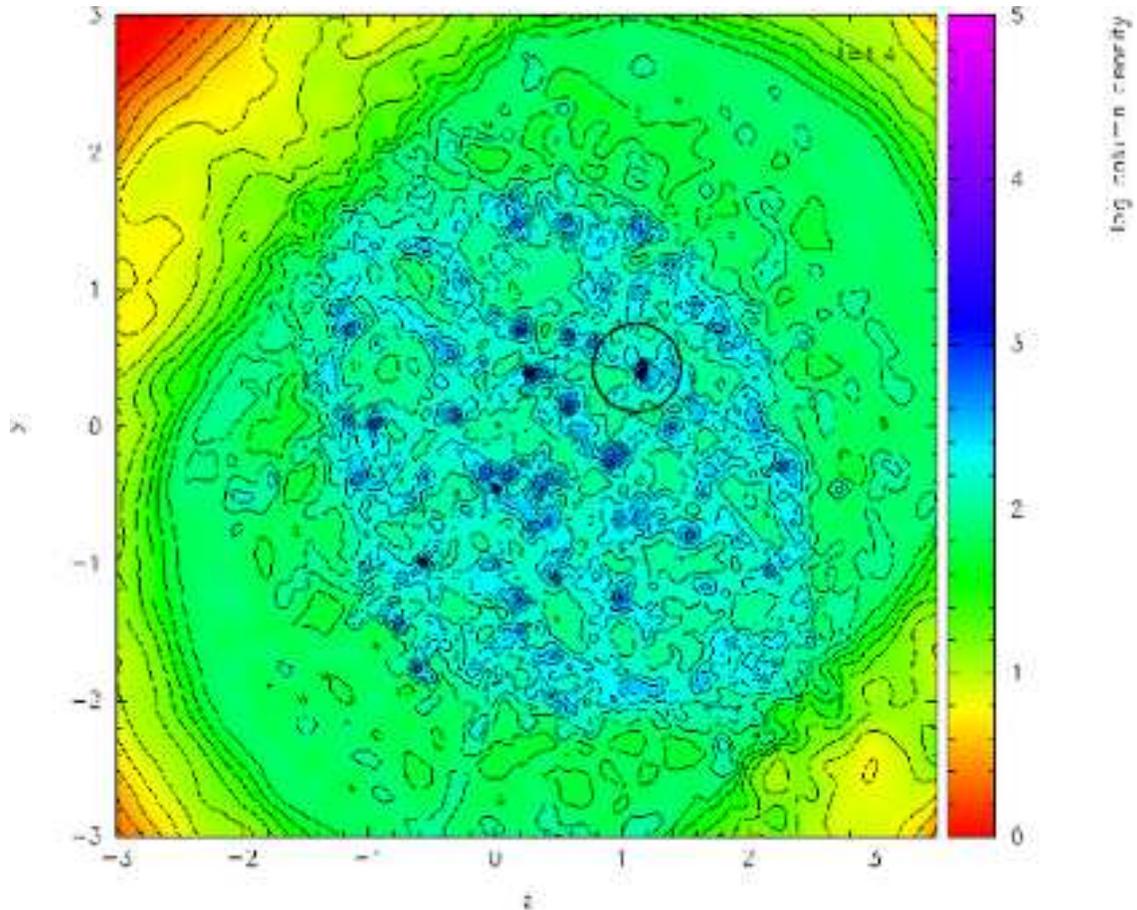}   
  \caption{A column density plot ($t$ = 1.4 Myrs) showing a close view of the oblique slab in model 4. Fragmentation of the slab leads to formation of number of clumps and elongated structures as can be seen with the aid of density contours overlayed on the column density plot. The circled region shows a fragmenting clump.}\vfil} \label{landfig}
\end{figure*}
%

\begin{figure*}
\centering
\vbox to 130mm{\vfil 
     \includegraphics[angle=270, width=13.0cm]{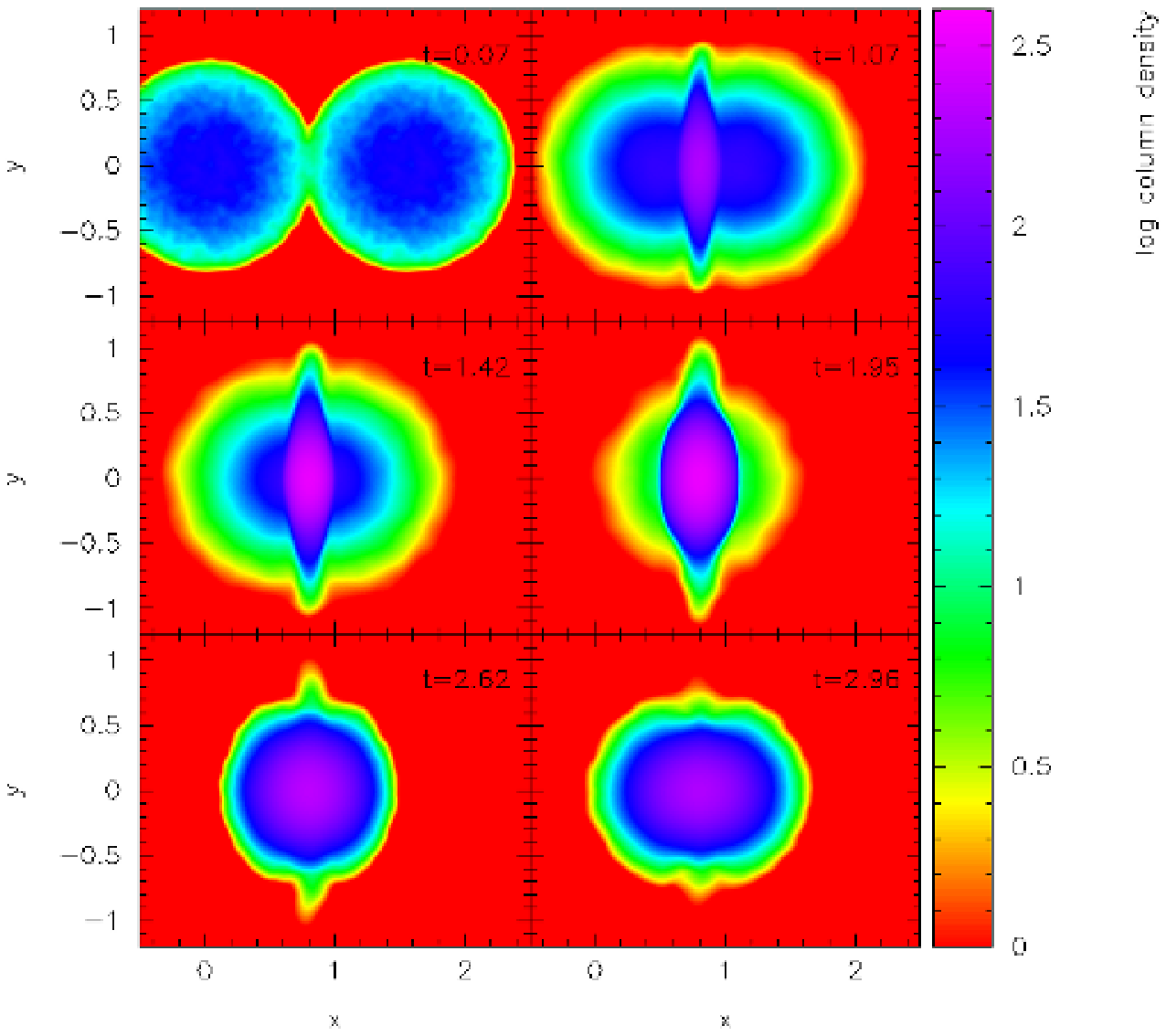}   
  \caption{A time (measured in Myrs) sequence of column density plots in model 5. Collision of clouds, followed by the formation of a prolate gas cloud ($t$ = 1.07 Myrs) and then its re-expansion is evident from these plots. The expansion of the post-collision gas cloud terminates in a diffuse cloud ($t$ = 2.96 Myrs). Material squirting from the top and bottom ends (\emph{jets}) of the slab are also visible in these plots. } \vfil} \label{landfig}
\end{figure*}

\textbf{Slow cloud collision}
  
Finally, we examine the case of a very low velocity head-on cloud collision. As discussed above, an adiabatic EOS, given by equation (2) with $\gamma = \frac{5}{3}$ is used to  model the weak shock in this case. Fig. 8 shows a time sequence of column density plots of the colliding clouds. Post-collision, the clouds merge and form a prolate gas cloud ($t$ = 1.42 Myrs in Fig. 8). The cloud is thermally supported against self gravity. The thermal pressure gradient within the cloud is non-isotropic due to its spheroidal shape.

Post-collision, as material from the colliding clouds streams in to the gas cloud, it progressively becomes denser and therefore warmer, as dictated by the stiff EOS employed in this model.  This builds up a thermal reservoir within it, and the cloud, therefore starts re-expanding along the collision axis ($t$ = 1.96 Myrs, $t$ = 2.62 Myr and $t$ = 2.96 Myrs in Fig. 8). During this process it flattens, and the re-expansion  continues till the thermal pressure along the $y$-axis, arrests the lateral collapse. This observation in this simulation is in consonance with the qualitative predictions of Ramsay (1961) and Mestel (1965). Note that, the cloud is not self gravitating and the lateral collapse is just a phenomenon involving gravo-thermal balance.

From the snapshots in Fig. 8, we can also see  that some material is ejected, also called \emph{jets}, from the top and bottom ends of the cloud. The qualitative feature of this model is similar to that of models 1 and 2 discussed above. However, the post collision gas cloud evolves on a much larger timescale than in cases 1 and 2. \emph{This is yet another scenario that could lead to the formation of diffuse clouds in the ISM. Star formation can commence in such clouds  only when they become sufficiently dense to support gravitational instability.}


\section{Conclusions}

 Colliding clouds dissipate kinetic energy and produce gas slabs which may fragment, leading to formation of density structures of various shapes and sizes. Some of these density concentrations may collapse and form stars while others may suffer tidal disruption. This model does not require injection of turbulence to trigger star formation and, stellar feedback from one episode of star formation may lead to other such events in the neighbourhood.

Numerical simulations by Clarke \& Gittins (2006) for instance, have shown that a burst event creates local perturbations in the galactic disk which in turn generate spiral patterns in it. Interference of many such perturbations may create complex structure, that might evolve on the timescale of a rotation period. The evolution of a galaxy essentially depends on the global star formation rate. Star burst events also play an important role in the chemistry of the intercloud medium. Powerful winds from young star clusters drive shock fronts and, while the abundance of molecular species may suffer attrition following the post shock ionisation, the pool of heavier elements will be replenished via stellar evolution. Cloud collision is a prominent mechanism of triggered star formation  and supposedly, also plays a pivotal role in the distribution of prestellar core masses and stellar population in the ISM (McKee 1999; Tan 2000).

In the present work, having explored the paradigm of low velocity cloud collisions, we have also tried to ascertain the importance post-shock cooling albeit in a rather crude manner, upon the dynamical evolution of the pressure confined slab. Our simple investigation shows that, under suitable conditions, gas slabs become Jeans unstable and fragment to produce clumps and  filaments. We conclude that colder slabs are more likely to fragment otherwise, the slab simply re-expands and ends up as a diffuse cloud in dynamical equilibrium. The former scenario is more interesting from the perspective of star formation, either multiples or larger $N$-body clusters. Density perturbations in the gas slab grow purely from white noise and therefore do not require any external trigger. In the latter scenario however, the diffuse gas clouds may become self gravitating, when crushed by sufficient external pressure else, clouds of various shapes simply lie scattered in the ISM. 

Models 1, 2, and 5 in the present work, belong to this latter paradigm while models 3 and 4, to the former. Figs. 4 and 7 respectively, show the fragmented gas  slabs in models 3 and 4. Clumps and well defined, elongated density structures are evident in both. Indeed, a few clumps in either slabs, have become self gravitating by the time respective calculations were terminated. The collapse of individual clumps however, could not be followed further due to shortage of resources. However, fragmentation of gas slabs is evidently  a propitious mode of clump formation. Clumps may eventually spawn stars and larger clouds may become wombs for larger star clusters.
 
The results of our off-centre cloud collision experiments (both models 2 and 4) differ significantly from similar experiments performed by Lattanzio \emph{et al} (1985), Chapman \emph{et al} (1992) and Bhattal \emph{et al} (1998) among others. These authors report a bar mode fragmentation of the post-collision oblique pressure compressed slab, which is  different from the fragmentation observed in our model. On the contrary, we observe that the oblique gas slab either, simply re-expands to finally produce a diffuse gas cloud (model 2) or,  fragments gravitationally to form clumps and filaments (model 4). We attribute this difference to the fact that the mass of the slab which, in our simulation is at least an order of magnitude greater than that in the work cited above. As a result of larger mass and therefore higher column density, the length of the unstable mode is shortened and grows much faster.

 We admit that the investigation presented here is too simplified and dwells purely on a gravo-thermal treatment of the problem. At best, we have a skewed picture before us for a more elaborate future study.  For an elaborate study of the subject, magnetic field needs to be included and post-shock radiative cooling needs to be better treated (\emph{c. f.} V{\'a}zquez-Semadeni \emph{et al} 2007). Also, collapsing clumps can be followed further by employing a radiative transfer scheme (\emph{c. f.} Stamatellos \emph{et al} 2007), rather than a simple barytropic EOS used here.

\begin{acknowledgements}
    This work was completed as part of post graduate research and funded through a studentship (DSW/Edu/Inf/2004/6283(35)) awarded by the State Government Of Maharashtra, India. All the column density plots presented here were prepared using the publicly available graphics package, SPLASH prepared by Dr. D. Price (Price 2007). 
Dr. S. V. Anathpindika specially thanks Dr. Simon Goodwin for providing the latest version of his SPH code, DRAGON. Useful comments and suggestions by an anonymous referee are greatly appreciated.

\end{acknowledgements}

\end{document}